# Cyclone preparedness strategies for regional power transmission systems in data-scarce coastal regions of India

## Author Information

Affiliations

**Indian Institute of Technology Gandhinagar, Gandhinagar, India**

Surender V. Raj, Udit Bhatia & Manish Kumar

## Contributions

Conceived and designed the study: S.V.R., M.K., and U.B., collected data: S.V.R., and M.K., performed numerical simulations: S.V.R., analyzed and interpreted the results: S.V.R., and M.K., developed the figures: S.V.R., and U.B., wrote the draft manuscript: S.V.R., and M.K., contributed to the writing of the manuscript: U.B.

## Corresponding author

Correspondence to: Manish Kumar

## Abstract

As frequency and intensity of tropical cyclones, and degree of urbanization increase, a systematic strengthening of power transmission networks in the coastal regions becomes imperative. An effective strategy for the same can be to strengthen select transmission towers, which requires consideration of network at holistic scale and its orientation relative to coastline, towers' fragilities, and cyclones' properties. Since necessary information is often missing, actionable frameworks for the prioritization remain elusive. Based on publicly available data, we assess efficacies of strategic interventions in the network that serves 40 million people. After evaluating 72 strategies for prioritization, we find that strategies that consider rather simplistic properties of the network and its orientation with respect to the coastline work much better than those based purely on the network's properties, in spite of minor variations in towers' fragilities. This integrated approach opens avenues for actionable engineering and policy interventions in resource-constrained and data-deprived settings.

## Introduction

Increasing population density in the coastal regions of developing and developed nations would translate to enhanced exposure of supporting infrastructure[1], including dense and interconnected power transmission, water distribution, transportation, and communication networks, to "grey swan" tropical cyclones (i.e., high impact storms with non-negligible probabilities)[2–6]. With the projected increase in intensity and frequency of tropical cyclones due to continued rise in global mean surface temperatures, temperature gradients, and atmospheric moisture[7,8], strategic capacity enhancement of critical infrastructure is essential to mitigate the short- and long-term consequences of such events[9–12].

Augmented risk to power transmission systems due to tropical cyclones has gained significant attention from researchers and policymakers alike[13]. These systems' performance directly impacts the functioning of critical infrastructure systems, including lifeline networks (e.g., transportation, communication, water, and wastewater systems). A power transmission network is a multi-component system comprising of towers and substations. Towers support the conductor wires between substations, while substations supply power to the neighborhood. There has been a considerable interest to understand the network response and recovery characteristics of power transmission and distribution systems both from structural and infrastructural perspectives[14–19]. For example, Winkler et al.[14] performed a comparative assessment of multiple power transmission networks' performances in the event of cyclones. This was one of the earliest studies, wherein fragilities of transmission towers and substations were explicitly considered. Thereafter, multiple studies have focused on developing and evaluating frameworks for quantification of networks' resilience, i.e., the ability of a system to come back to its state before a disruption[15–17]. Furthermore, a recent research integrated the power transmission and water distribution networks, and studied the interdependence in performance of the system in the aftermath of a cyclone[20]. We note that while the aforementioned studies have focussed on highly localized networks (e.g., less than 5,000 sq. km.), many cyclone events in the past have impacted regions spanning hundreds of kilometres, thus offering limited or no actionable insights to infrastructure managers and stakeholders towards real-world recovery efforts (see Fig. 1.a). Further, previous studies are either based on synthetic data or make context-specific assumptions. Hence, their generalizability to the regions outside the study areas remains unclear. An exception to the aforementioned studies is by Pantelli et al.[17], where a representative large-scale power transmission network was considered. Authors evaluated alternate strategies to strengthen the transmission network and noted that strengthening the transmission corridors was the most effective way of improving the resilience of the network. We note that a practicable time- and cost-effective strategy to prioritize the corridors (and towers) for strengthening would need to consider sufficiently fine details of the network and its orientation with respect to the coastline, realistic properties of cyclones before and after making a landfall, and realistic fragilities associated with the transmission towers. We further argue that for implementation in real-world scenarios (where crucial data on properties of network, cyclones and fragilities is sparingly available, and there is considerable uncertainty associated with the available information), the parameter to establish the sequence of strengthening of towers should be effective and simple to determine. We emphasize that such prioritization would be particularly useful in reducing the financial and societal impacts in the aftermath of a cyclone, especially for large administrative units (e.g., a state spawning in the area of hundreds of thousands of square kilometers) with limited financial resources.

In this research, we develop a systematic approach to identify the prioritization strategy for strengthening (or hardening) power transmission towers, which is demonstrated through a real-world large-scale power transmission network in the region of Odisha, India. The power transmission network comprises 41,000+ transmission towers spanning over an area of 155,000 km$^2$ covering a coastline approximately 500 km long, and serving over 40 million people. The state has been hit by major cyclones in recent decades, causing devastating loss of life and property[4,21,22]. Annual per capita income of Odisha is approximately INR 100,000 (~USD 1,400), which places the state among the poorest in India[23]. We collected data on towers damaged during Cyclone Fani (2019) from Odisha Power Transmission Corporation Limited (OPTCL) through a Right to Information (RTI) Act application. Wind speeds at the locations of towers were estimated based on data available from Indian Meteorological Department (IMD) and a radial wind speed model[24]. We developed fragility curves for transmission towers and studied their efficacy. We then generated a series of *Fani-like* cyclone tracks that could hit the coast of Odisha. A total of 72 strategies to identify transmission towers for strengthening were evaluated. These strategies considered different geographical regions in which towers were to be hardened, methods to prioritize towers for hardening within a geographical region and numbers of towers to be hardened, and/or extents of hardening. We found that some intuitive strategies could be most effective in reducing the loss of functionality, while hardening the towers might not yield meaningful gains even if substantial number of towers are strengthened in sub-optimal ways.

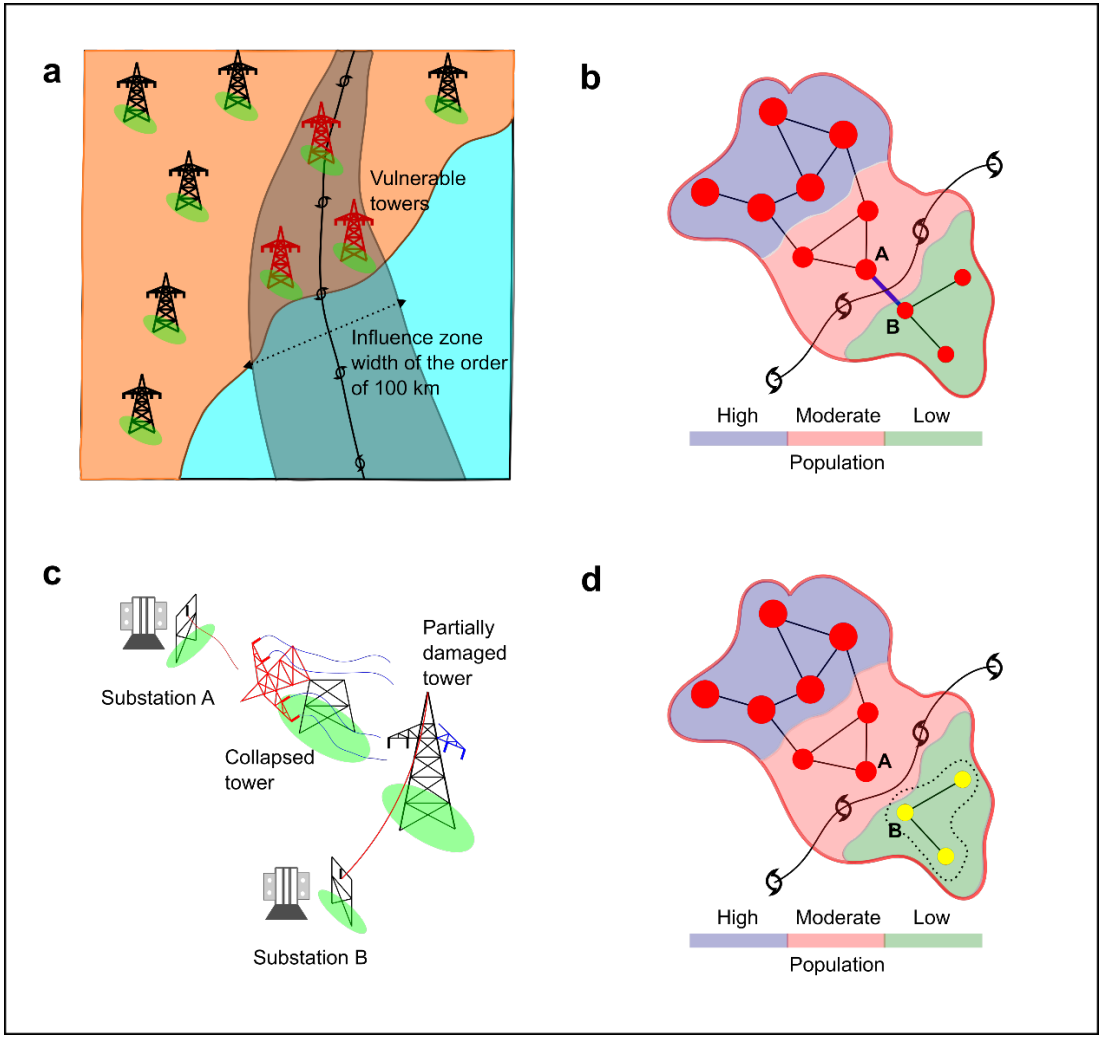

**Fig. 1: Effect of tropical cyclones on the power transmission network**

**a.** A representative cyclone scenario with its trajectory and zone of influence (not to scale) indicated. Transmission towers falling in the zone (shown in red) may sustain structural damages. **b.** A representative power transmission network serving a region with varying population density. Red dots indicate the substations sized based on the population served and the black lines represent transmission corridors. Trajectory of a cyclone passing through a transmission corridor between substations A and B is also shown. **c.** Transmission towers in corridor A-B suffer structural damages due to wind forces leading to disruption in power supply. **d.** Power transmission network is divided into two segments as corridor A-B is dysfunctional. Substations in the smaller segment of the network are shown in yellow. The associated population will not receive power supply immediately after the cyclone.

## Results

Fragility functions for transmission towers in Odisha

We develop fragility curves for high-voltage transmission towers (e.g., 132 kV, 220 kV) in Odisha based on the damage observed during Cyclone Fani. Partial damage (e.g., damage near the peak of the tower or in a cross-arm) was observed in 41 towers, while 87 towers had collapsed. Wind speeds at the locations of the transmission towers are estimated using the Willoughby double exponential model[24] for tropical cyclones (see Methods). Locations of these towers and contours of wind speeds are shown in Fig. 2.a. Two possible damage states for a transmission tower are considered: (1) structural collapse, and (2) functionality disruption. A total of 87 and 128 towers experienced the two damage states during Cyclone Fani, respectively. Fragility curves are developed using the Bounding Engineering Demand Parameter (EDP) method (see Methods)[25,26] for the two damage states (see Fig. 2.b). These curves compare reasonably well with the fragility curves reported from other parts of the world[17,27,28], as can be seen from Supplementary Figure 1.

We evaluate the efficacy of the functionality disruption fragility curve next. The network is subjected to the simulated Cyclone Fani 1,000 times[15]. The random number-based method for assigning a damage state (see Methods) is used to determine if a tower is in the said damage state. Fig. 2.c shows the number of towers in the damage state for each of the 1,000 simulations. The average number of towers in the damage state is 132, which compares well with the number of towers damaged during Cyclone Fani (= 128). A comparison of locations of damaged towers obtained through a simulation and that observed during Cyclone Fani is made in Fig. 2.d.

**Fig. 2: Characterizing fragility of high-voltage transmission towers in the state of Odisha, India**

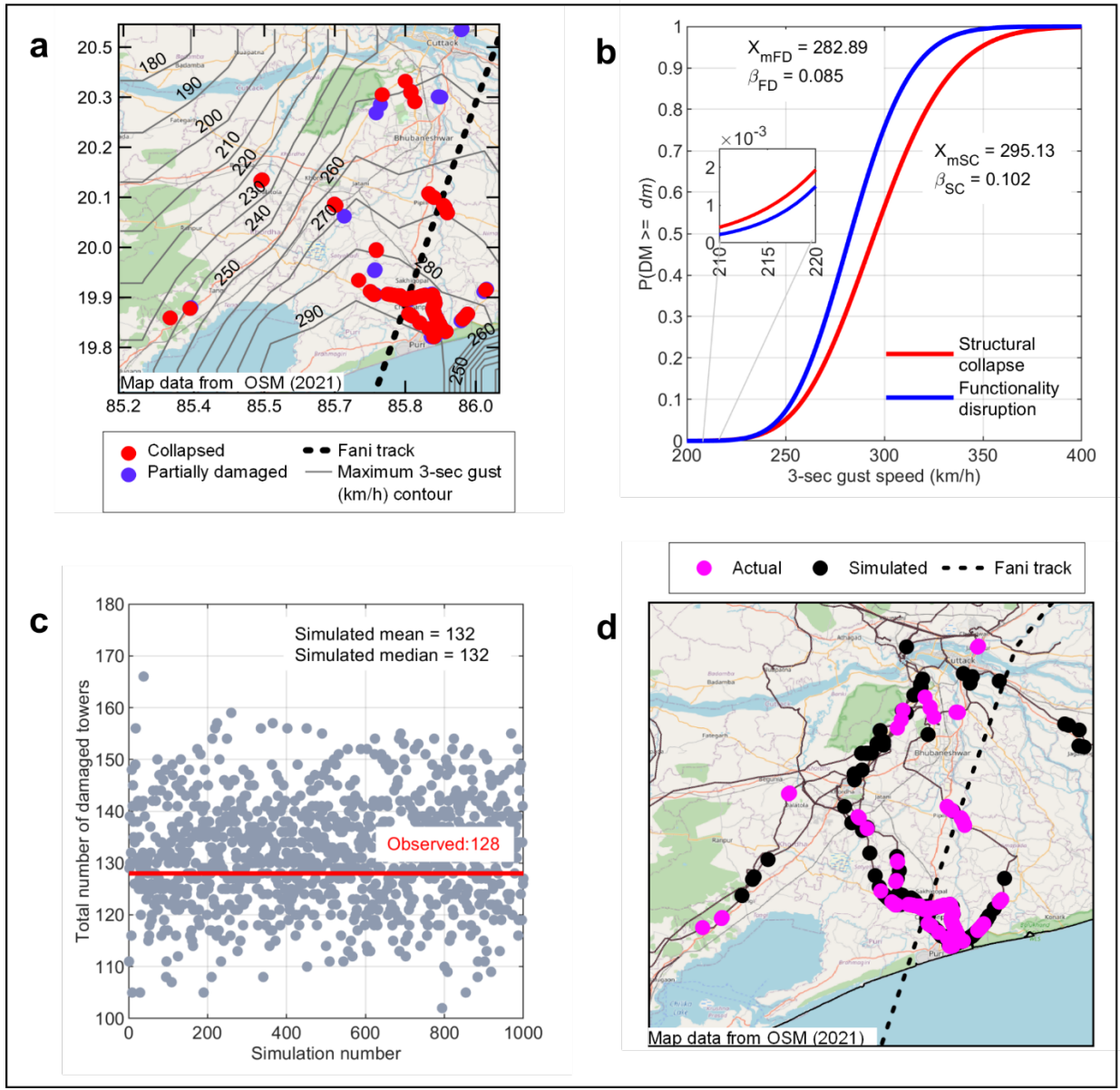

**a.** Locations of collapsed and partially damaged transmission towers (132 kV and 220 kV) during 2019 Cyclone Fani along with contours of maximum 3-sec gust speed. The track of cyclone is also shown. **b.** Fragility curves for transmission towers in the state of Odisha for two damage states (structural collapse and functionality disruption) based on the damage observed during the 2019 cyclone. **c.** Number of towers in the functionality disruption damage state for 1,000 simulations of Cyclone Fani. **d.** Locations of towers in the functionality disruption damage state for a simulated Cyclone Fani scenario and of the towers observed to be in the damage state during 2019 Cyclone Fani.

## Response of Odisha's power transmission network to *Fani-like* scenarios

We generated *Fani-like* cyclone tracks by displacing and rotating the observed track of 2019 Cyclone Fani (see Methods). Cyclones making a landfall between latitudes 17.12°N – 21.77°N only could damage a transmission tower in Odisha (see Supplementary Figure 2). The pattern of damage to towers was considered to be captured consistently when approximately 3,000 *Fani-like* tracks were included in the study (see Supplementary Figure 3). The network was subjected to each track 1,000 times[15] (see Supplementary Figure 4) and average response (e.g., number of damaged towers) was considered. Fig. 3.a shows the number of damaged towers (averaged over 1,000 runs) for the 2,882 *Fani-like* scenarios. Expectedly, more towers are damaged for a track if there are more towers in the vicinity of the landfall. It is worth noting that the trajectory of cyclone can significantly influence the number of damaged towers. As an example, the number of damaged towers corresponding to landfalls at 21.42°N and 21.46°N are 256 and 540, respectively. These locations are rather close to each other and the large difference in the number of damaged towers can be explained by the difference in path of cyclones along the coast before landfall. Fig. 3.b presents an example of how a cyclone can damage towers before making landfall. Distance between the farthest damaged towers was 83 km (corresponds to Fig. 3.b) and 230 km for the two landfall locations, respectively (see Fig. 3.a). These lengths are of the order of or are considerably greater than the dimensions of realistic power transmission networks studied in the past[15,29], which underscores the need to study these networks at a sufficiently large scale. A composite effect of landfall location and path length is presented in Fig. 3.c.

**Fig. 3: Response of Odisha's power transmission network to *Fani-like* scenarios**

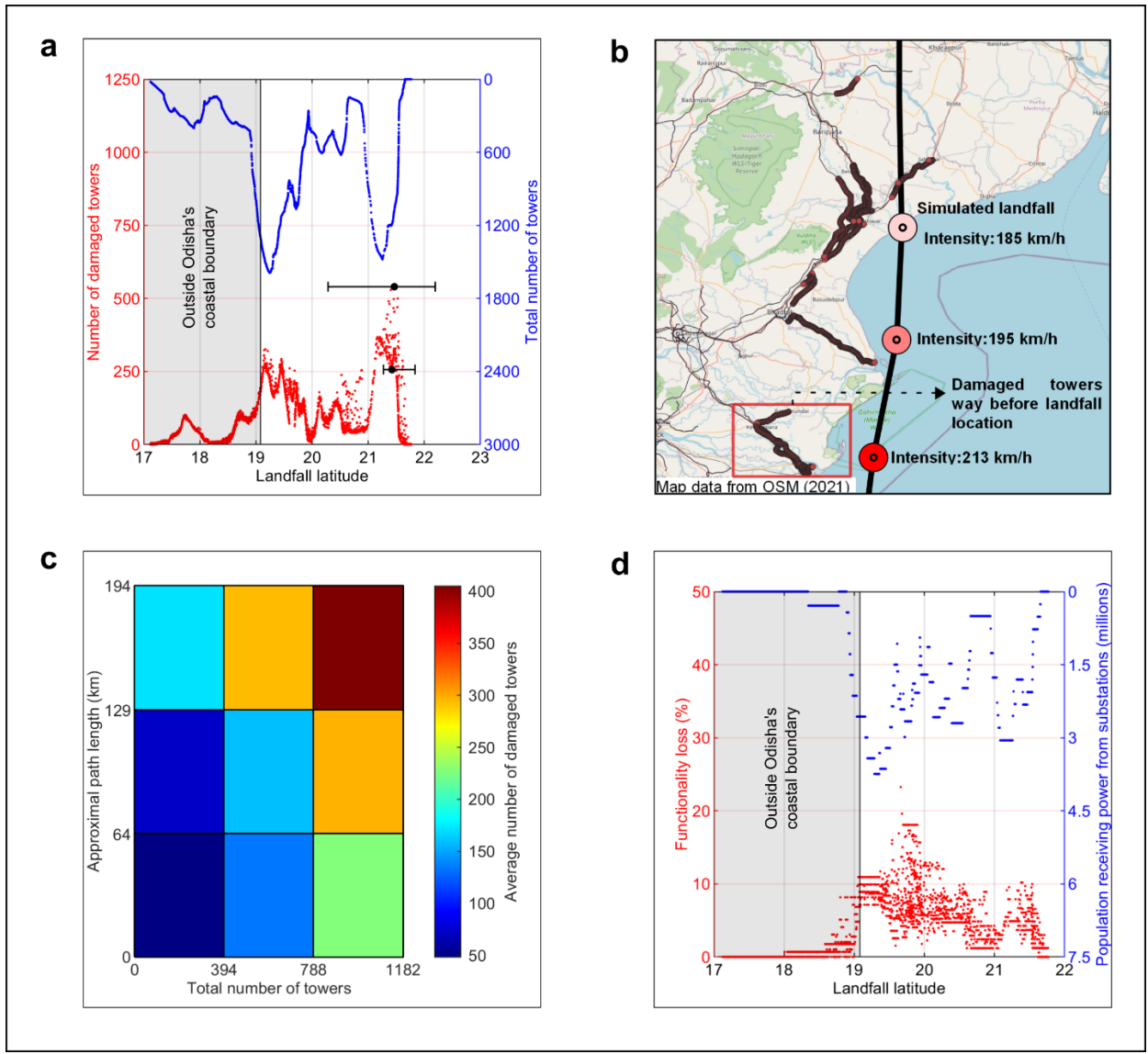

**a.** Number of damaged towers (left vertical axis) and total number of towers within the 50 km radius from a landfall location (right vertical axis) plotted against latitude for corresponding landfall location. Ranges of latitudes of damaged towers for two cyclone scenarios are also identified using horizontal black lines. **b.** A cyclone scenario with the path traversed close to the coast before landfall **c.** Average number of damaged towers for different combinations of number of towers within 50 km radius of a landfall location and length of path traversed by the cyclone within 50 km from the coast before making landfall **d.** Functionality loss in terms of percentage of population without power (left vertical axis) and the population receiving power supply from substations within 50 km radius of the landfall location (right vertical axis) plotted against the landfall latitude.

Functionality loss for the network (see Methods) is defined as the percentage of the state's population not receiving electricity after a Cyclone[19,29]. Functionality loss correlates rather moderately with the number of towers in the vicinity (50 km radius) of the landfall (correlation coefficient = 0.63; see Supplementary Figure 5a). The correlation with the population receiving power supply from the substations (see Methods) within 50 km radius of the landfall is relatively better (correlation coefficient = 0.79; see Fig. 3.d and Supplementary Figure 5b). This is expected since damage to one or many towers in a transmission corridor can disrupt the power supply to the population associated with the corresponding substations. Population associated with a corridor (or substation) can be used as a basis to prioritize towers for hardening.

## Alternate strategies to strengthen Odisha's power transmission network

We evaluate 72 strategies to reduce the functionality loss in the transmission network. These strategies differ in the choice of geographical regions where towers are to be strengthened, methods to prioritize the towers for hardening within a geographical region, number of towers to be hardened within a geographical region, and extent of hardening a tower. Since the goal of this study is to understand the functionality loss due to cyclones, only the geographical regions close to the coast of Odisha are considered. The first geographical region (IS) includes the area corresponding to the basic wind speed of 180 km/h per Indian standard IS 875, Part III[30] (see Fig. 4.a). The second region (RD) includes four segments of radius 50 km, which are picked intuitively to cover major clusters of substations near the coast (see Fig. 4.b). The number of towers in the two regions are 10,995 and 4,229, respectively. The third geographical region (CB) is identified in a way that it has approximately 3,000 towers and is associated with an optimal functionality loss (see Methods and Fig. 4.c). Total area under the three geographical region are 43,111 km², 16,818 km² and 31,413 km², respectively. These regions cover 10 – 28% of the area and 8 – 26% of the towers in the state of Odisha.

**Fig. 4: Geographical regions in which transmission towers are to be strengthened**

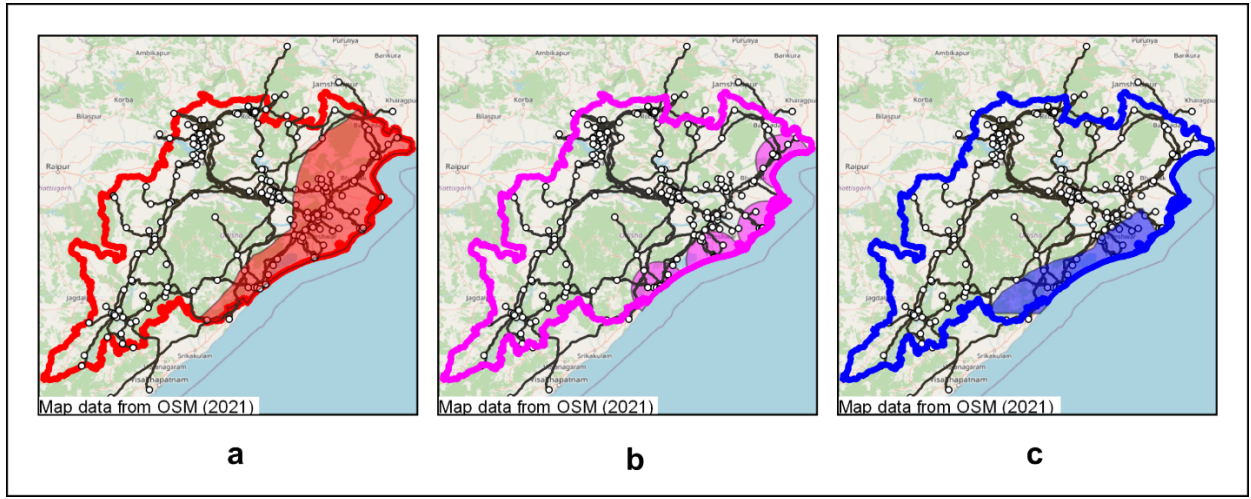

**a.** Area with basic wind speed of 180 km/h according to the Indian standard (IS)[30] **b.** Four patches of 50 km radius (RD) covering clusters of substations near the coast of Odisha. **c.** Coastal band (CB) with 3,345 towers and an optimal functionality loss (see Methods).

The four approaches to prioritize the towers to be hardened are based on their nearest distance to coast (ND) (see Fig. 5.a), the population weight associated with the transmission corridor (PW) (see Fig. 5.b), edge betweenness centrality (EB.)[31] (see Fig. 5.c), and functionality improvement index (FII) (see Fig. 5.d). Approaches based on PW and FII are described in Methods. Two values for number of towers to be strengthened are considered: 1,500 and 3,000. Three levels of hardening are considered, wherein the fragility curve corresponding to functionality disruption damage state (see Fig. 2.b) is shifted such that the wind speed corresponding to the 50% cumulative damage probability is increased by 10%, 20% and 100% keeping the associated logarithmic standard deviation unchanged (see Supplementary Figure 6).

**Fig. 5: Approaches to prioritize transmission towers for hardening in a geographical region**

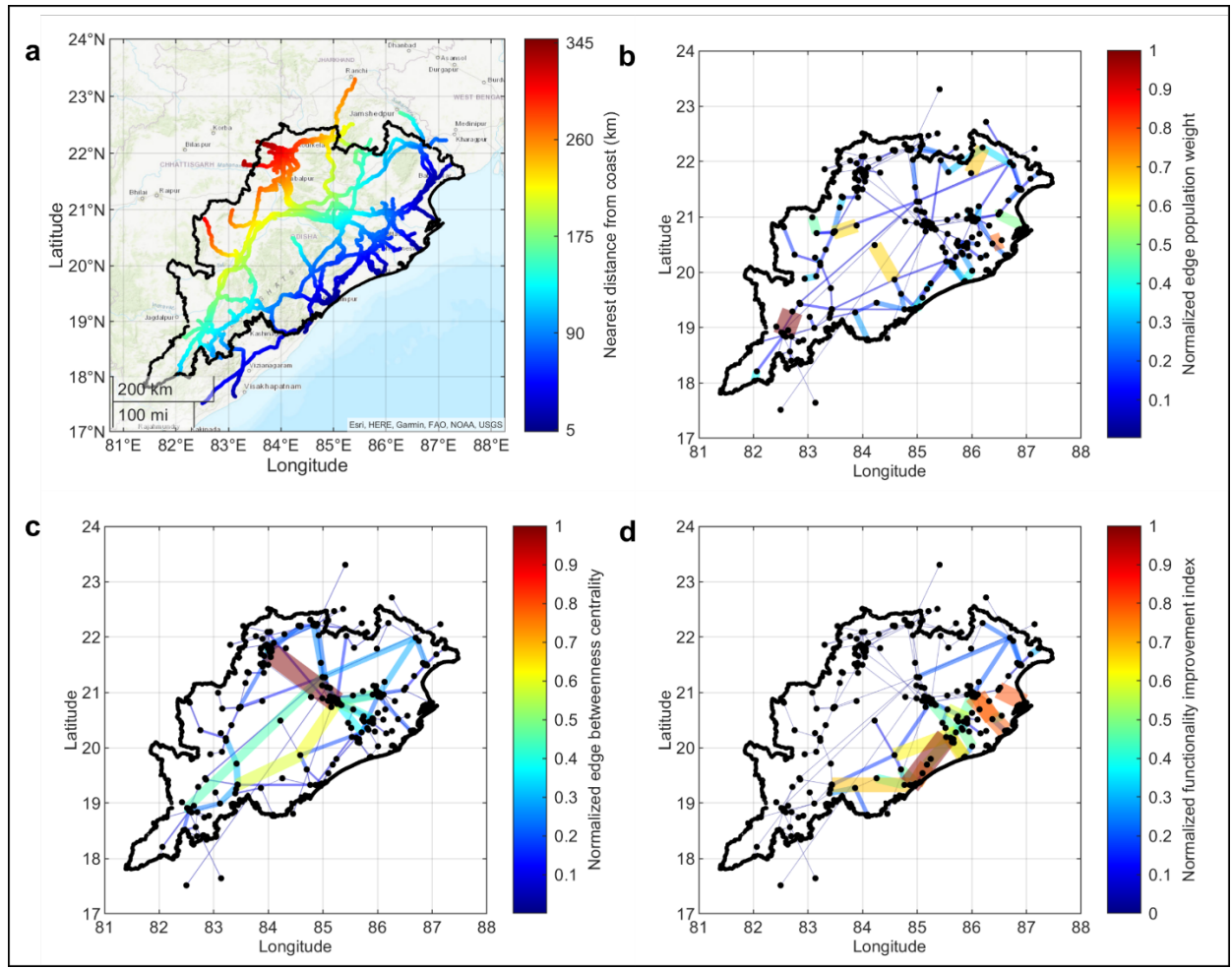

**a.** Odisha's power transmission network with the transmission towers identified using colours based on their nearest distance (ND) from the coast. **b. c.** and **d.** show idealised Odisha's power transmission network with the transmission corridors (edges) identified using colours (and size) based on their population weight (PW), edge betweenness centrality (EB) and functionality improvement index (FII), respectively.

The influence of 72 strategies to strengthen the towers is quantified in terms of two parameters: (1) average functionality loss, i.e., average percentage population of state that experienced disruption in power supply due to cyclone (see Methods), and (2) average number of towers in the functionality disruption damage state (see Fig. 2.b). For each strategy, the network is subjected to two sets of approximately 3,000 tracks (see Supplementary Figure 2). The results for the two sets of tracks are presented in Supplementary Table 1 and Supplementary Table 2.

Prioritizing towers for hardening based on PW led to maximum reduction in the functionality loss (see Fig. 6.a). This is expected since the functionality loss considered herein is a direct measure of the population affected during a cyclone. The efficacy of the criteria was particularly high when geographical regions RD or IS (in that order) were considered, when 3,000 towers were strengthened, and/or extent of hardening was greater. The criteria based on FII was found second-most effective. Parameter FII accounts for the population as well as the response of network subjected to a cyclone scenario, and may be more useful when other measures to define functionality are considered.

Selection of towers for hardening through the ND approach led to least number of towers being damaged for all combinations of geographical region, number of towers hardened and extent of hardening. The approach was most effective when geographical region IS or RD (in that order) was considered (see Fig. 6.b). The above observations suggest that the choice of the geographical region for strengthening the network can be based on design standards or intuition.

The approach to prioritize towers for hardening based on EB criteria was least efficient in reducing the functionality loss or the number of damaged towers (see Fig. 6.a and Fig. 6.b), which is expected because this measure tends to prioritize towers that are located "centrally" in a region rather than those near the coast. This approach was chosen because the focus of this study is on strengthening the relevant transmission corridors (substations are not considered affected directly by cyclonic winds[17,32]) and this is one of the few network science-based parameters to enable the same.

The geographical region CB (see Fig. 4.b) corresponds to a patch with an optimal set (see Methods) of approximately 3,000 towers. However, strengthening of towers in this region leads to an inferior outcome compared to that in other two regions, which highlights the need to consider the power transmission system holistically for the purpose of strengthening.

**Fig. 6: Alternate strategies evaluated based on mean total damaged towers and mean functionality loss**

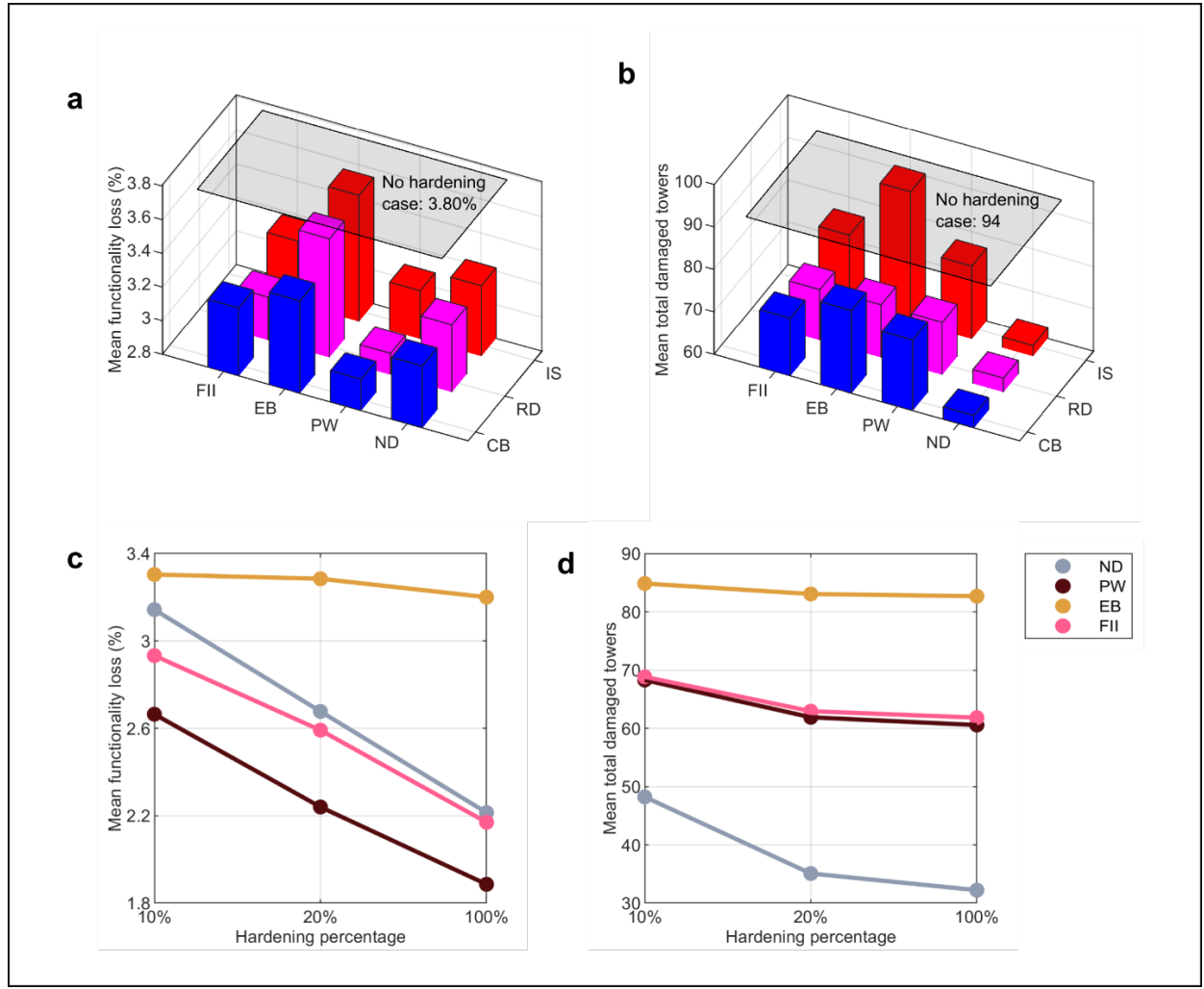

**a.** and **b.** show the functionality loss, i.e., population not receiving the power supply after a cyclone, and total number of damaged towers averaged over 2,882 cyclone scenarios, respectively. These results correspond to 1,500 towers hardened by 20%. Also indicated in the panels are average functionality loss and average number of damaged towers when no tower was hardened. **c.** and **d.** present the average functionality loss and number of towers damaged, respectively, for different levels of hardening the towers achieved for the four approaches to prioritize the towers for hardening. These results are for geographical region I.S. and 3,000 towers hardened. Also indicated in the two panels are the average functionality loss and average number of damaged towers when no tower was hardened.

Hardening of towers is effective only when a suitable strategy for prioritization of towers is considered. As an example, hardening the towers does not improve the functionality or help reduce the number of towers damaged materially if approach EB is considered (see Fig. 6.c and Fig. 6.d). Increasing the extent of hardening can lead to a reduction in the functionality loss, i.e., population affected (see Fig. 6.d). However, advantages are rather limited if the goal is to reduce the number of towers damaged (see Fig. 6c).

## Discussion

With increasing urbanization and global warming (and associated changes in the climate), it is expected that the establishments in the coastal regions will be further exposed to the tropical cyclones. These cyclones impact the infrastructure causing massive economic losses and downtime. Strategic strengthening of the vulnerable infrastructure is therefore necessary. Such an exercise would require studies performed at a reasonably large scale and under assumptions relevant to the context. However, lack of necessary data often hinders these efforts. We study the power transmission network in a coastal state of India with a population of over 40 million and coastline longer than 500 km. This state has among the lowest per capita income in India, and has been battered by major tropical cyclones in recent decades. Limited data on properties of transmission towers, wind speeds during previous cyclones, damages observed etc. exists in the public domain. We obtained relevant data on the location of over 40,000 transmission towers in the state, and on the 128 towers that received significant damages during 2019 cyclone Fani. We also got data on maximum wind speeds during the Cyclone Fani from publicly available sources. We used an existing radial wind speed model to estimate wind speeds at the locations of the tower. Subsequently, we developed fragility curves for the transmission towers. We studied the response of transmission network in the state under simulated *Fani-like* cyclone scenarios. Number of damaged towers in a given cyclone scenario depends on the number of towers in the region close to the landfall and on the proximity of cyclone track to the coastline prior to the landfall. A cyclone can damage towers that are 250 km or more away, which highlights the need to study a power transmission network at a relatively large scale. We defined functionality loss as the population not receiving the power supply in the aftermath of a cyclone. Expectedly, the functionality loss correlated more strongly with the population associated with substations compared to number of towers in the vicinity of the landfall.

We considered 72 alternate strategies to prioritize transmission towers for hardening. The first choice was made for the geographical region in which towers are to be hardened. Within a region, the towers were prioritized based on intuitive, simulation-based and network science-based parameters. Different number of towers and different levels of hardening were also considered. We found that a simplistic selection of geographical region (e.g., region of high wind speed per the relevant design standard, clusters of substations rather close to the coast) can lead to effective reduction in the functionality loss. Within a geographical region, the methods based on intuition and simulation led to much improvement in functionality compared to network science-based parameters, which is expected because the latter parameters would prioritize towers that are often far away from the coast. This observation highlights the possibility of designing a transmission network for improved performance during cyclones. Expectedly, the greater number of towers strengthened or a greater level of strengthening led to a better functionality. However, meaningful improvements were achieved only when efficient strategies for prioritizing the towers was used. The conclusions presented above did not change considerably if hardening from 0% to 10% or from 10% to 20% was considered. Fragility curve considered in the present study did not include the effect of variations and uncertainties in the estimation of wind speeds, terrain conditions etc. Such consideration would alter the fragility curves. Estimated direct loss during Cyclone Fani was INR 24,000 crores (USD 3.34 billion). Losses to electrical transmission and distributions systems constituted approximately 30% of this loss. Assuming that the cost of strengthening a tower by 20% is similar to that of restoration of damaged towers (e.g., INR 35,00,000 (or USD 48200) as noted in the RTI response), the cost of strengthening 3,000 towers would be approximately INR 1,000 crores (or USD 140 millions), which is much less compared to the direct losses incurred during the cyclone. Such strengthening could mean that approximately 1 million less population is affected during a cyclone.

## Methods

## Data and pre-processing

All data presented in this paper are from publicly available sources. Geographical boundaries of the Indian state of Odisha and 30 districts in the state are extracted from the district-level shapefile available at the ArcGIS Hub (https://hub.arcgis.com). The district-wise population data are obtained from the 2011 census of India (https://censusindia.gov.in). Latitudes and longitudes of the 41,814 transmission towers within the boundaries of the state of Odisha are taken from OpenStreetMap (O.S.M.) dataset (https://openstreetmap.org). These towers are identified in Supplementary Figure 7. The locations of 128 damaged towers and the extent of damage in these towers are collected from Odisha Power Transmission Corporation Limited (OPTCL) through a Right to Information (R.T.I.) Act application. The locations of these towers are shown in Fig 2.a. We collected the data of track for cyclone Fani from the Indian Meteorological Department (IMD) (http://www.rsmcnewdelhi.imd.gov.in).

The geographical locations of the transmission towers are imported in the open source geographic information system software platform QGIS. The locations of the ends of the transmission corridors are identified using the O.S.M. Standard basemap. These locations are recognized as substations and are considered as nodes of the idealized transmission network for Odisha (see Supplementary Figure 7). An edge between two nodes is created if there exists at least one transmission corridor connecting the two nodes. For the purpose of further computations, an adjacency matrix $A(i, j)$ is created which stores number of edges between nodes $i$ and $j$. The size of the matrix is equal to number of nodes. The power transmission network consists of 227 nodes and 282 edges.

## Determination of 3-sec gust speed at a location

The *stormwindmodel* package of R language (https://cran.r-project.org/web/packages/stormwindmodel, accessed June 2020) provides the details of steps involved in calculating the wind speed at the location of towers. This package imports a cyclone track and desired grid points (i.e., locations where wind speeds are required) as inputs and calculates the wind speed using Willoughby double exponential radial wind profile model for cyclones[24]. Steps involved in the calculation of wind speed are executed in the numerical computing software MATLAB.

Indian Meteorological Department reported 3-min maximum sustained wind speed during 2019 Cyclone Fani. Relevant Indian design standards[30] present the design wind speeds in terms of 3-sec gust speed. The latter definition of wind speed is considered in the present study. The wind speed reported by IMD is multiplied by 1.58[33] to obtain the 3-sec gust speed.

## Fragility curve

A fragility curve plots the probability of exceedance of a certain damage state (e.g., structural collapse) of a tower against the wind speed. The damage data obtained from OPTCL includes 87 towers that collapsed and 41 towers that received partial damage. The partial damage comprises damage to the cross-arms and to the peak of the tower. Such damages also imply disruption in the power supply. Bounding *edp* method[25,26] is used to develop the fragility curves corresponding to the two damage states: (1) structural collapse, and (2) functionality disruption.

Maximum 3-sec gust speeds at the locations of the 41,814 towers in the network are determined. The towers are grouped into 30 bins of identical size based on 3-sec gust speeds (first bin: 54.9 – 62.7 km/h; last bin: 281.7 – 289.5 km/h). The basis for the number of bins is presented in Supplementary Figure 8. Average wind speed is calculated for each bin. For each of the two damage states, the ratio of the number of towers in a particular damage state in the bin to the total number of towers in the bin is calculated. We then fit a lognormal cumulative distribution function into the data comprising ratio and average wind speed for the bins, which is considered as the fragility curve for this study and is described by the following expression:

$$F_{dm}(im) = \phi\left[\frac{\ln(im / X_{mdm})}{\beta_{dm}}\right] \qquad (1)$$

where, $F_{dm}$ represents the probability of exceedance of damage state $dm$, $im$ refers to the intensity measure, i.e., 3-sec gust speed, $X_{mdm}$ refers to the median and $\beta_{dm}$ refers to the logarithmic standard deviation of the fragility function corresponding to the damage state $dm$.

## State of functionality for a transmission tower

Following approaches are considered to determine if a tower is in a particular damage state in case the network is subjected to a cyclone.

1. Random number-based: For a given cyclone scenario, the maximum wind speeds at all towers are estimated. The probability of exceedance of a damage state is obtained from the fragility curve for a damage state. A uniformly distributed random number between 0 and 1 is generated for each tower. If the number is less than the probability of exceedance, then the tower is considered to be in the damage state, and vice-versa. This approach has also been considered in the past[15].

2. Damage probability group-based: Wind speeds are determined at the locations of towers for a given cyclone scenario. The probability of exceedance for a damage state is determined using the corresponding fragility curve. The towers are grouped into 10 bins based on the probabilities of exceedance. For example, a bin can have towers with the exceedance probabilities 0.3 – 0.4. The number of towers from each bin in the damage state is the product of the total number of towers in the bin and the average exceedance probability for the bin. The towers in the damage state are selected randomly from a bin. Since the state of Odisha has 41,814 towers with most of them falling in the exceedance probability bin 0.0 – 0.1. As a consequence, the number of towers in a particular damage state will be very high. Therefore, it was found necessary to determine a suitable lower bound for the bin with the lowest exceedance probabilities. This bound was varied between 0.0 and 0.1. The lower bound set equal to 0.0155 led to the total number of towers in the functionality disruption damage state in the network subjected to the 2019 Cyclone Fani equal to 128, which is same as that observed during the cyclone (see Supplementary Figure 9). Accordingly, the lower bound of the lowest exceedance probability bin was set equal to 0.0155. Since this approach is able to consistently capture the number of towers damaged, it is used in this paper to establish the damage state for a tower, unless specified otherwise.

## Population weight assignment to nodes and edges

We obtained the population of 30 districts in the state of Odisha from Census 2011[34]. The population density map is shown in Supplementary Figure 10. The population of a district is divided equally among all the substations within the boundary of the district. The population associated with a substation is referred to as population weight of the node. The nodes outside the geographical boundary of the state of Odisha are assigned a population weight of zero. The population weight for the edge joining two nodes is obtained using the following expression:

$$PW_e = \frac{PW_{n1}}{k_{n1}} + \frac{PW_{n2}}{k_{n2}} \qquad (2)$$

where, $PW_e$ is the population weight of the edge, $PW_{n1}$ and $PW_{n2}$ are the population weights of the nodes connected by the edge, and $k_{n1}$ and $k_{n2}$ are the number of edges connected to the two nodes, respectively.

## Loss of functionality for a transmission network

For a given cyclone scenario, the damage state for the towers can be determined as discussed in Methods. All the transmission corridors in which one or more towers are in 'functionality disruption' damage state are considered dysfunctional. An edge is removed from the network, if all the transmission corridors associated with the edge are dysfunctional. Subsequently, a node is considered to have lost functionality if it is out of the 'giant component'[35]. Loss of functionality in the power transmission network is defined as the percentage of state population without power supply[19,29], wherein the affected population is calculated as the sum of the population associated with the nodes out of the giant component. Mathematically, it is expressed as follows:

$$F_L = \frac{\sum_{i=1}^{N} P(i) \times x(i)}{\sum_{i=1}^{N} P(i)} \tag{3}$$

where, $i$ represents a node, $P(i)$ is the population weight of the $i^{th}$ node, and $x(i)$ is equal to 1 if the node is out of the giant component, and 0 otherwise. Functionality of a network $F_A$ after being subjected to a cyclone is given as follows:

$$F_A = 1 - F_L \tag{4}$$

## Functionality improvement index

Functionality of a network subjected to a *Fani-like* cyclone scenario is determined first per Eq. (5). Subsequently, an edge of the network is assumed to be hardened infinitely. The functionality of the network after hardening is determined for the same cyclone scenario. Functionality improvement index $FII$ for an edge is defined as follows:

$$FII = \frac{1}{N_S} \sum_{j=1}^{N_S} D_j \times \left[ \frac{F_{A,h}^{\ j} - F_{A,o}^{\ j}}{F_{A,o}^{\ j}} \right] \times 100 \tag{5}$$

where, $j$ represents a simulated cyclone scenario, $N_S$ is the number of cyclone scenarios, $D_j$ is equal to 1 if the edge is damaged during the cyclone scenario, and zero otherwise, $F_{A,o}^{\ j}$ represents the functionality of the original network for the cyclone scenario, and $F_{A,h}^{\ j}$ represents the functionality of the network after hardening the edge for the $j^{th}$ cyclone scenario. A similar index in the context of resilience has been used in the past[17,18].

## Cyclone track simulations

Data is available for a very limited number of cyclones that hit the state of Odisha in the past. In order to understand the influence of a cyclone on the network, a much larger dataset is needed. Accordingly, simulated tracks are generated[36] herein by shifting the landfall location of the 2019 Cyclone Fani (see Fig. 2.a) to random locations along the coast with its track rotated counter-clockwise at random angles between 0 and 90 degrees. Latin Hypercube Sampling technique[37] is considered to establish the combinations of landfall locations and angle of rotation for the simulated tracks.

The locations of landfall for a simulated track that can potentially damage the transmission towers within the state of Odisha are identified first. Landfall locations in the Bay of Bengal at latitudes ranging between 13.15°N (Chennai, India) and 22.35°N (Chittagong, Bangladesh) are considered. These latitudes cover a significant coastal boundary of Bangladesh, and the Indian states of Tamilnadu, Andhra Pradesh, Odisha and West Bengal. Three sets of 500 *Fani-like* cyclone tracks are generated that are considered to make the landfall in the specified range of latitudes. Landfall locations are identified for which a tower in the state of Odisha can be damaged for each set of tracks (See Supplementary Figure 2).

A total of 3,000 fresh *Fani-like* cyclone tracks are generated that make a landfall within the range of latitudes identified above. Some of these tracks make landfall at the coast of Odisha after making a landfall previously (see Supplementary Figure 11). While these scenarios are possible, it is recognized that the intensity of cyclones decreases considerably after a landfall[38]. Such tracks are not considered in the present study.

## Identification of coastal band (CB) geographical region

We attempted to identify a band along the coast of Odisha such that hardening the towers within the band would lead to an 'optimal' reduction in the loss of functionality. Width of the band was set equal to 100 km, which corresponds to the distance from the coast up to which the towers were observed to be damaged for the approximately 3,000 *Fani-like* scenarios (see Methods). A total of 83 bands were considered that differed in the length and location along the coast. Each patch had approximately 3,000 towers, which is the upper limit of number of towers considered for strengthening in the present study. Significance of each band was quantified as follows. Functionality loss corresponding to cyclone tracks making landfall within the band was set equal to 0.00%. Subsequently, the functionality loss averaged over approximately 3,000

*Fani-like* scenarios for the network was determined (see Supplementary Figure 12). The band associated with the maximum reduction in the functionality loss was considered as the geographic region of interest. The functionality loss corresponding to the original network was 3.80%, while that corresponding to the selected coastal band was 1.32%.

## Data Availability

All data reported in the paper has been obtained from publicly available sources identified in the paper. Complete data sets would be made available upon request to the corresponding author.

## Code Availability

The code necessary to generate information in this manuscript is available at https://github.com/surendervraj/Cyclone-prepardness-for-regional-power-transmission-network.

## Acknowledgements

Financial support provided by Indian Institute of Technology Gandhinagar, India is gratefully acknowledged.

## Ethics declarations

Competing interests

Authors declare no competing interests.

## Supplementary Information

A supplementary document is provided comprising figures and tables that support or complete some assertions made in the paper.

# Supplementry Information

## Supplementry Fig

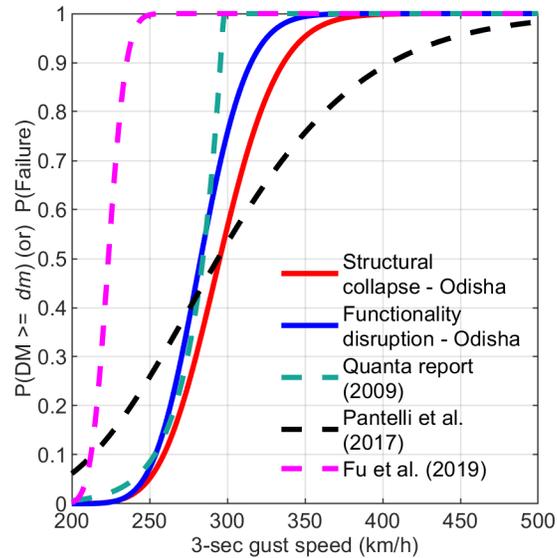

Supplemetry Figure 1: **Fragility functions for transmission towers in Odisha, India and in other parts of the world.** The fragility curves for Odisha compares well with that reported for towers in the United Kingdom[1]. The median for the fragility curves for transmission towers in the United States[2] compares well with those for Odisha, but the former is associated with a greater standard deviation. The fragility curves for the towers in China[3] appear similar to those for Odisha, except that these seem to correspond to towers in a region of lower wind speeds compared to those considered in the present study.

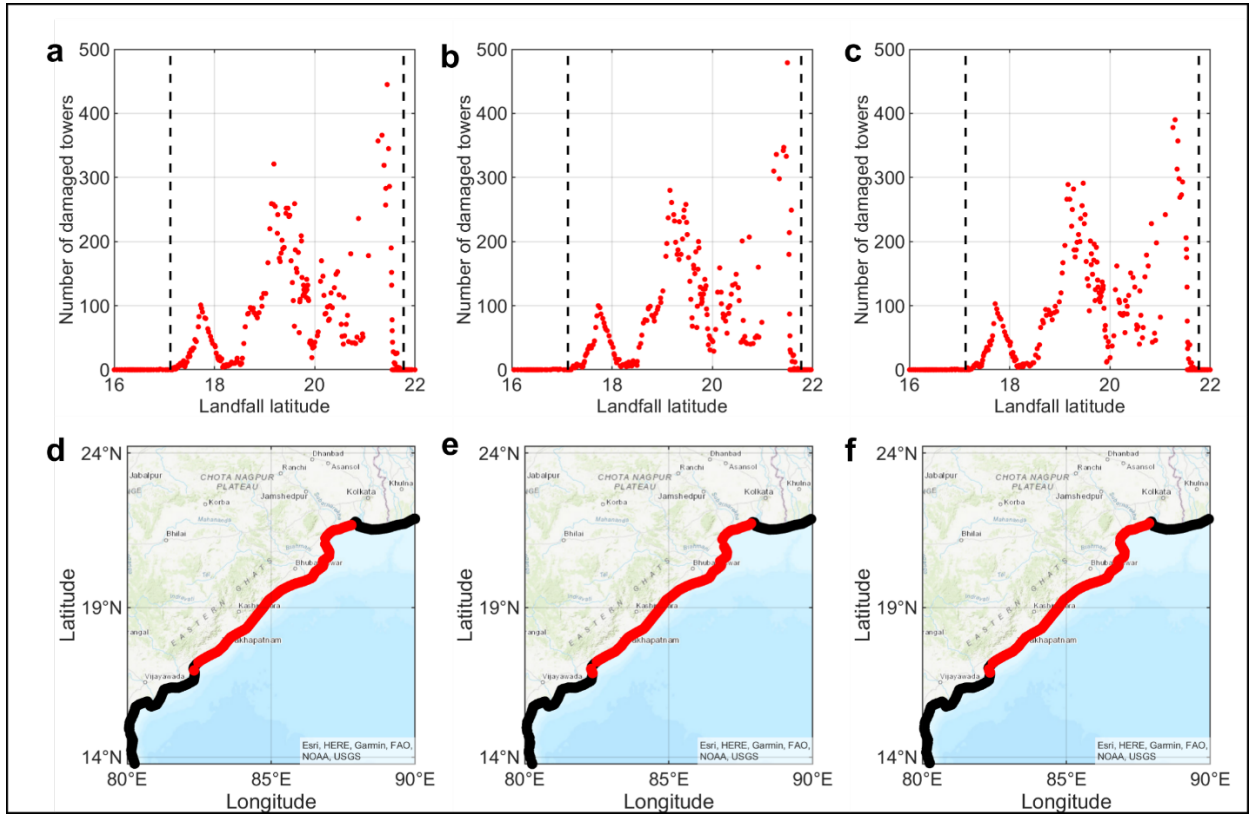

Supplemetry Figure 2: **Locations of landfall for *Fani-like* scenarios causing damage to transmission towers in the state of Odisha. a, b** and **c** each show the plots of total number of damaged towers against the corresponding landfall latitude of three separete sets of simulated 500 *Fani-like* tracks. These tracks had a landfall between Chennai (India) and Chittagong (Bangladesh), covering the entire coastal length of Odisha (India). The broken vertical lines indicate the boundaries outside which no simulated *Fani-like* track making landfall caused damage to any tower in the state of Odisha. **d, e** and **f** each show the eastern coast between Chennai and Chittagong highlighted. The segment highlighted using red corresponds to landfall locations of the *Fani-like* tracks that caused damage to at least one tower in Odisha.

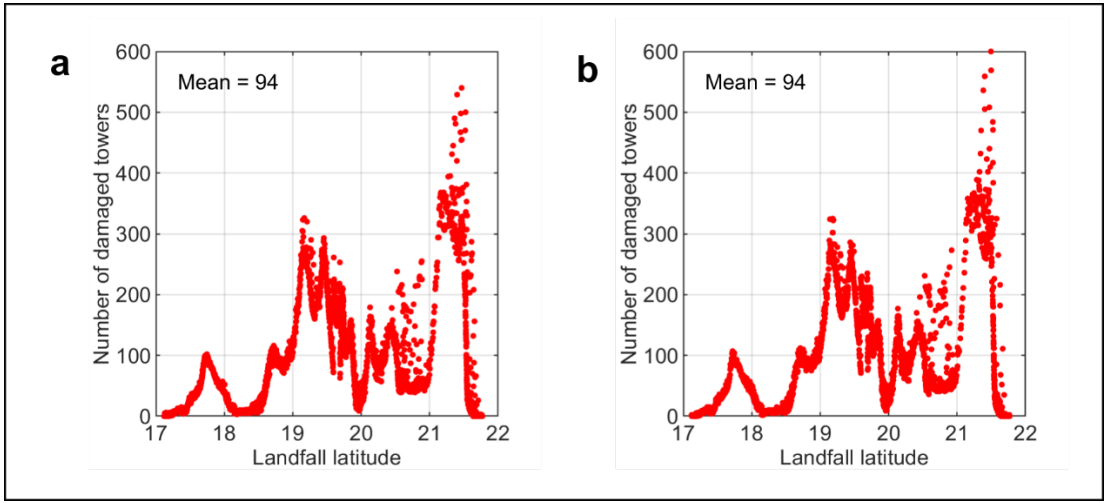

Supplemetry Figure 3: **Response of Odisha's power transmission network under two sets of *Fani-like* scenarios. a.** Number of damaged towers corresponding to a landfall latitude for 2,882 *Fani-like* cyclone tracks generated through shifting of the landfall location and rotation of the Fani track observed during 2019. **b.** Number of damaged towers corresponding to a landfall latitude for 2,876 *Fani-like* cyclone track. Approximately, 3,000 tracks can lead to a consistent pattern of damage in transmission towers.

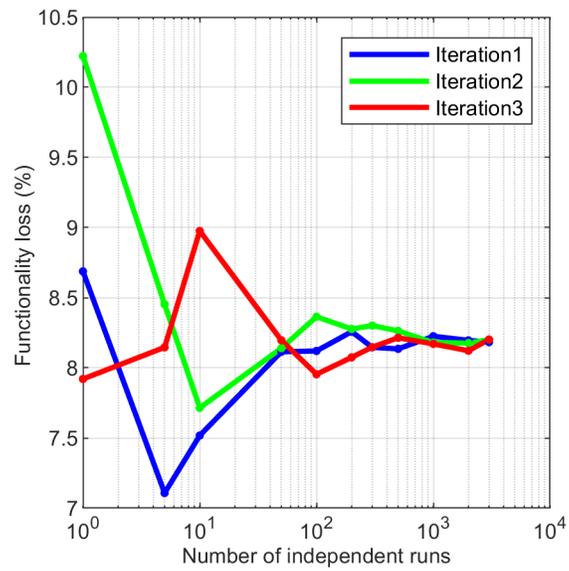

Supplemetry Figure 4**: Functionality loss estimates averaged over the number of times the power transmission network of the state of Odisha is subjected to the 2019 Cyclone Fani.** It is seen that the estimate of average functionality loss converges for number of runs 1,000 or greater (also see Methods for assigning a damage state).

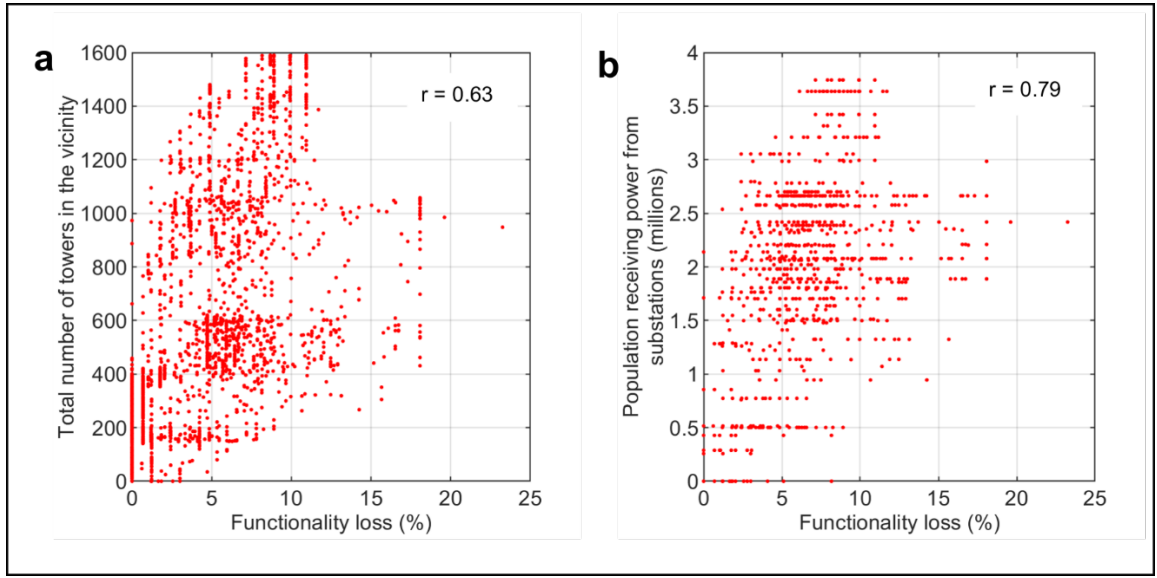

Supplemetry Figure 5: **Dependence of functionality loss corresponding to a cyclone scenario on the parameters of the power transmission network in the vicinity of the landfall. a.** Total number of tower within 50 km radius of the landfall location plotted against functionality loss, i.e., percentage of state population not receiving the power supply in the immediate aftermath of cyclone (correlation coefficient = 0.60). **b.** Total population associated with the substations within the 50 km radius of the landfall location plotted against functionality loss (correlation coefficient = 0.75).

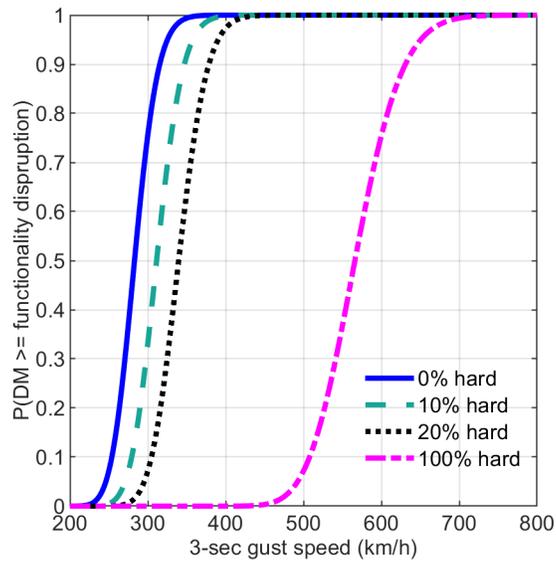

Supplemetry Figure 6: **Fragility curves corresponding to functionality disruption damage state in the state of Odisha.** Three levels of hardening a tower are considered, wherein the wind speed corresponding to 50% cumulative damage is increased by 10%, 20% and 100%, keeping the associated logarithmic standard deviation constant.

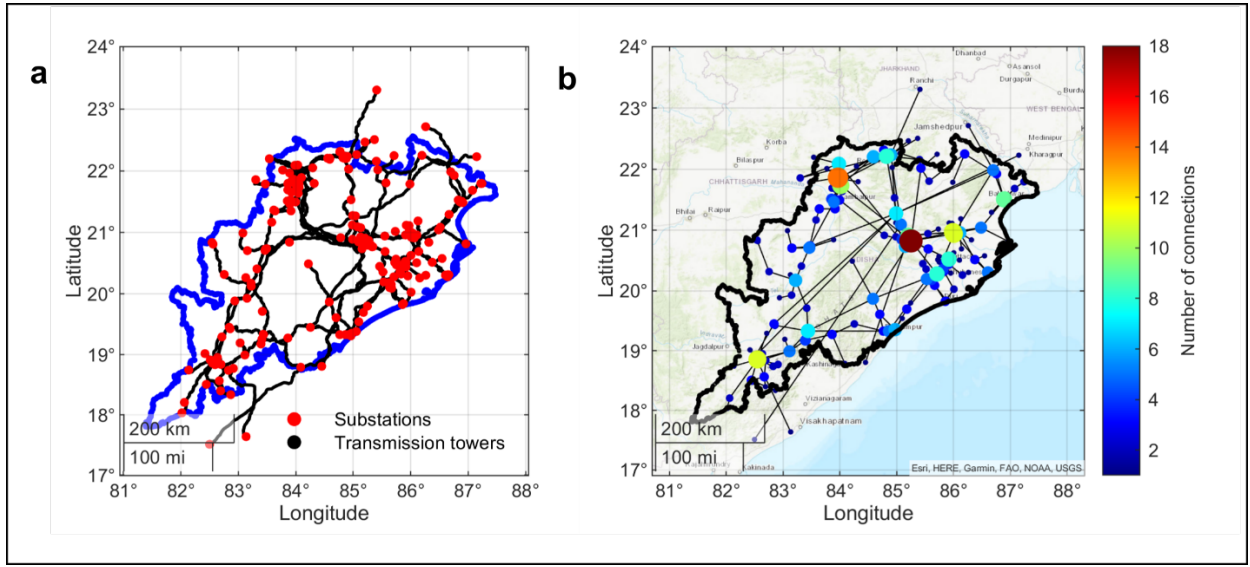

Supplemetry Figure 7: **Odisha's power transmission network. a.** Geographical locations of the 227 substations and 41,814 transmission towers serving the state of Odisha, India. The data is obtained from the Open Street Maps (OSM) database. **b.** Idealized power transmission network for the state of Odisha. Solid circles represent a substation idealized as a node, and the straight lines represent a transmission corridor. The substations are colored and sized based on the number of connections with the neighboring substations (i.e., degree).

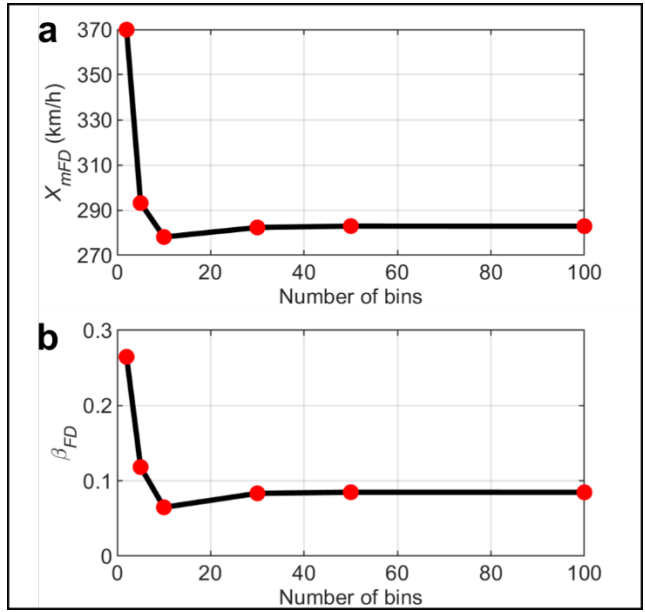

Supplemetry Figure 8: **Variation of fragility curve parameters with number of bins. a.** and **b.** show median of functionality disruption damage state fragility function and associated logarithmic standard deviation plotted against number of bins, respectively. The fragility curve parameters converge for number of bins 30 or greater.

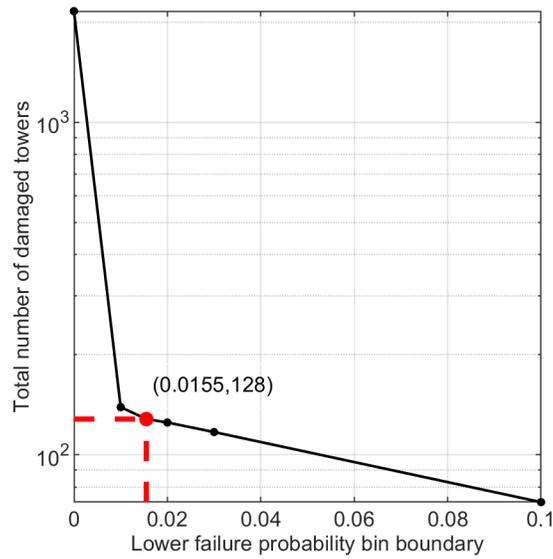

Supplemetry Figure 9: **Determination of lower bin boundary to predict damage by failure probability binning.** The process of predicting the damage by binning the failure probabilities is repeated for different values of lower bin boundary for the actual track of Cyclone Fani. It is observed that the lower bin boundary chosen as 0.0155 predicts the exact number of damaged towers (128) during Cyclone Fani. This value is chosen as the lower bin boundary for all the simulations in this study.

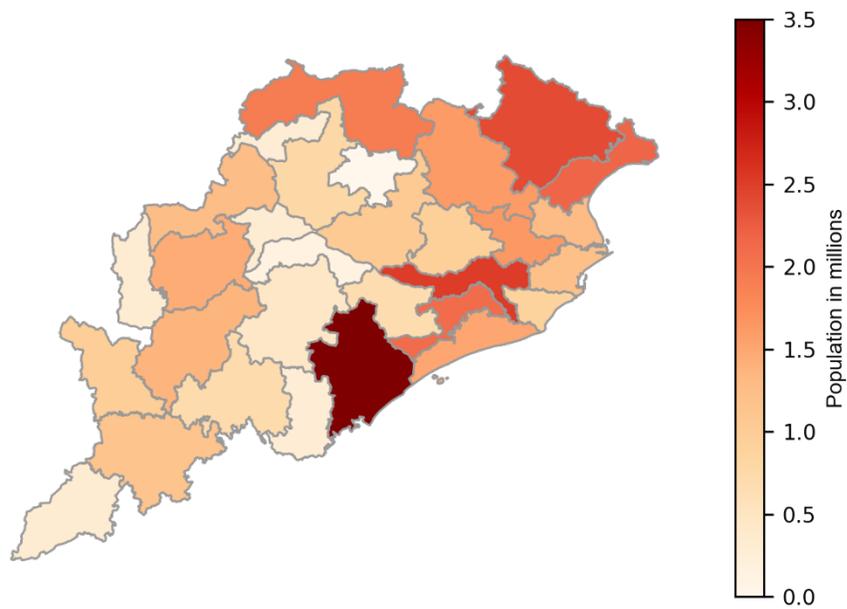

Supplemetry Figure 10: **District-wise population map of the state of Odisha, India based on 2011 Census** [4]

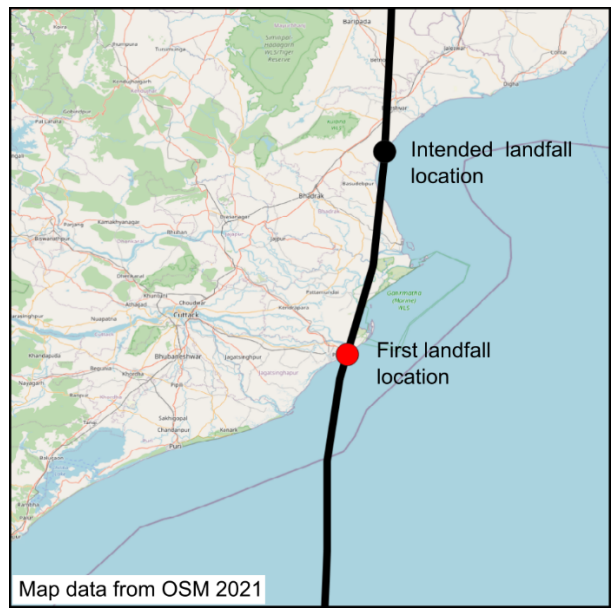

Supplemetry Figure 11: **An example of a simulated track making a landfall before the intended location of the landfall.**

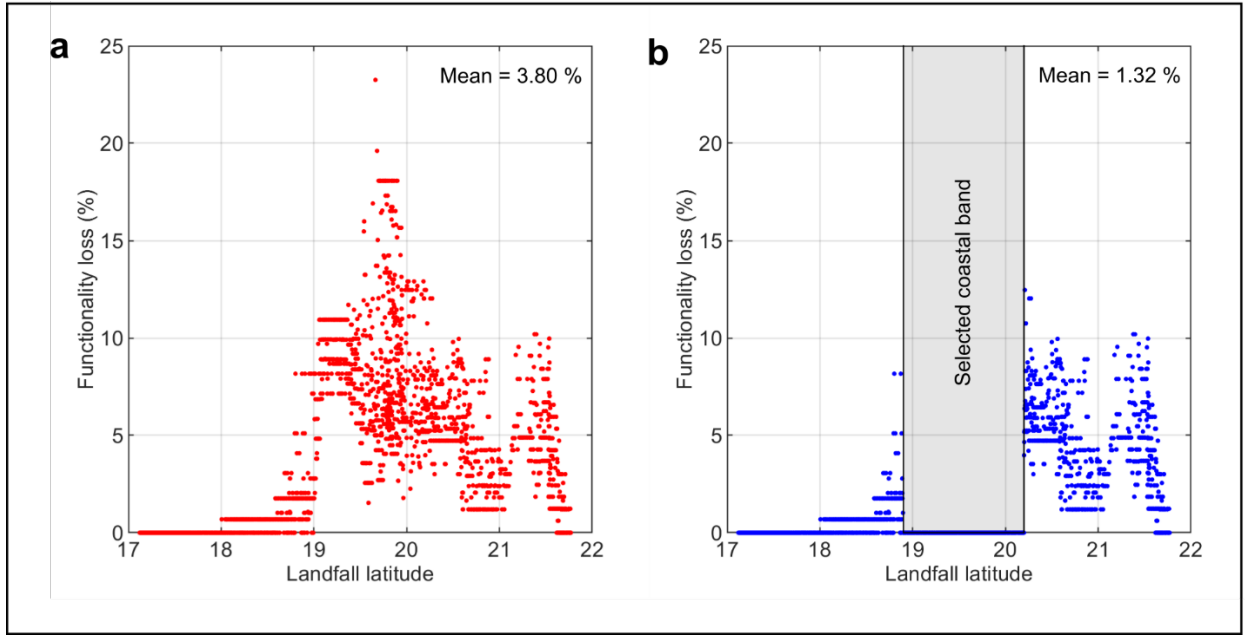

Supplemetry Figure 12: **Selection of coastal band associated with an 'optimal' reduction in the loss of functionality. a.** Functionality loss (i.e., percentage of population without power supply after a cyclone) plotted against landfall latitude. These results are for the first set of approximately 3,000 tracks (see Supplementary Figure 3). **b.** Functionality loss plotted against landfall latitude with the functionality loss set equal to 0.00% corresponding to landfall locations within the selected coastal band.

Supplementary Tables

Supplementary Table 1: Functionality loss, i.e., percentage of population in the state of Odisha not receiving power supply after a cyclonic event, and number of towers damaged averaged over 2,882 *Fani-like* cyclone tracks as a function of choice of geographical region (IS, RD, CB), approach to prioritize towers for hardening (ND, PW, EB, FII), number of towers to be hardened (1500, 3000) and extent of hardening (10%, 20%, 100% increase in median wind speed). The average functionality loss and average number of damaged towers was 3.80% and 94, respectively. Smaller values in the table indicate a greater effectiveness of the strategy for hardening.

| Sl No | Region | Approach | Number of towers | Functionality loss (% state population affected) | | | Number of damaged towers | | |
|---|---|---|---|---|---|---|---|---|---|
| | | | | Level of hardening | | | Level of hardening | | |
| | | | | 10% | 20% | 100% | 10% | 20% | 100% |
| 1 | IS | ND | 1500 | 3.40 | 3.22 | 3.03 | 70 | 63 | 61 |
| 2 | IS | ND | 3000 | 3.14 | 2.68 | 2.21 | 48 | 35 | 32 |
| 3 | IS | PW | 1500 | 3.31 | 3.09 | 2.86 | 81 | 77 | 76 |
| 4 | IS | PW | 3000 | 2.66 | 2.24 | 1.89 | 68 | 62 | 61 |
| 5 | IS | EB | 1500 | 3.52 | 3.55 | 3.53 | 91 | 91 | 90 |
| 6 | IS | EB | 3000 | 3.30 | 3.28 | 3.20 | 85 | 83 | 83 |
| 7 | IS | FII | 1500 | 3.34 | 3.18 | 2.99 | 80 | 76 | 76 |
| 8 | IS | FII | 3000 | 2.93 | 2.59 | 2.17 | 69 | 63 | 62 |
| 9 | RD | ND | 1500 | 3.40 | 3.20 | 3.05 | 70 | 63 | 62 |
| 10 | RD | ND | 3000 | 3.12 | 2.65 | 2.17 | 51 | 38 | 36 |
| 11 | RD | PW | 1500 | 3.24 | 2.93 | 2.67 | 77 | 72 | 71 |
| 12 | RD | PW | 3000 | 2.85 | 2.22 | 1.61 | 59 | 50 | 48 |
| 13 | RD | EB | 1500 | 3.62 | 3.50 | 3.35 | 77 | 72 | 71 |
| 14 | RD | EB | 3000 | 3.05 | 2.62 | 2.03 | 60 | 51 | 49 |
| 15 | RD | FII | 1500 | 3.37 | 3.06 | 2.74 | 76 | 72 | 71 |
| 16 | RD | FII | 3000 | 3.03 | 2.53 | 2.06 | 60 | 51 | 49 |
| 17 | CB | ND | 1500 | 3.40 | 3.17 | 2.80 | 70 | 63 | 61 |
| 18 | CB | ND | 3000 | 3.19 | 2.60 | 2.15 | 64 | 56 | 55 |
| 19 | CB | PW | 1500 | 3.33 | 2.99 | 2.68 | 80 | 77 | 76 |
| 20 | CB | PW | 3000 | 3.13 | 2.54 | 2.14 | 66 | 59 | 58 |
| 21 | CB | EB | 1500 | 3.51 | 3.35 | 3.06 | 82 | 79 | 79 |
| 22 | CB | EB | 3000 | 3.19 | 2.72 | 2.33 | 68 | 61 | 60 |
| 23 | CB | FII | 1500 | 3.55 | 3.20 | 2.93 | 78 | 74 | 73 |
| 24 | CB | FII | 3000 | 3.09 | 2.56 | 2.10 | 67 | 60 | 59 |

Supplementary Table 2: Functionality loss, i.e., percentage of population in the state of Odisha not receiving power supply after a cyclonic event, and number of towers damaged averaged over 2,876 *Fani-like* cyclone tracks as a function of choice of geographical region (IS, RD, CB), approach to prioritize towers for hardening (ND, PW, EB, FII), number of towers to be hardened (1500, 3000) and extent of hardening (10%, 20%, 100% increase in median wind speed). The average functionality loss and average number of damaged towers was 3.75% and 94, respectively. Smaller values in the table indicate a greater effectiveness of the strategy for hardening.

| Sl No | Region | Approach | Number of towers | Functionality loss (% state population affected) | | | Number of damaged towers | | |
|---|---|---|---|---|---|---|---|---|---|
| | | | | Level of hardening | | | Level of hardening | | |
| | | | | 10% | 20% | 100% | 10% | 20% | 100% |
| 1 | IS | ND | 1500 | 3.35 | 3.18 | 3.01 | 70 | 63 | 61 |
| 2 | IS | ND | 3000 | 3.10 | 2.65 | 2.13 | 48 | 35 | 32 |
| 3 | IS | PW | 1500 | 3.24 | 3.05 | 2.84 | 80 | 77 | 76 |
| 4 | IS | PW | 3000 | 2.60 | 2.18 | 1.85 | 68 | 62 | 60 |
| 5 | IS | EB | 1500 | 3.47 | 3.44 | 3.47 | 91 | 90 | 90 |
| 6 | IS | EB | 3000 | 3.25 | 3.24 | 3.16 | 84 | 83 | 82 |
| 7 | IS | FII | 1500 | 3.30 | 3.17 | 2.96 | 79 | 76 | 75 |
| 8 | IS | FII | 3000 | 2.90 | 2.56 | 2.12 | 69 | 63 | 62 |
| 9 | RD | ND | 1500 | 3.41 | 3.17 | 3.01 | 70 | 63 | 62 |
| 10 | RD | ND | 3000 | 3.07 | 2.57 | 2.12 | 50 | 38 | 36 |
| 11 | RD | PW | 1500 | 3.21 | 2.87 | 2.59 | 76 | 72 | 71 |
| 12 | RD | PW | 3000 | 2.80 | 2.19 | 1.57 | 59 | 50 | 48 |
| 13 | RD | EB | 1500 | 3.58 | 3.43 | 3.25 | 77 | 72 | 71 |
| 14 | RD | EB | 3000 | 3.03 | 2.60 | 1.96 | 60 | 51 | 49 |
| 15 | RD | FII | 1500 | 3.33 | 3.00 | 2.72 | 76 | 71 | 71 |
| 16 | RD | FII | 3000 | 2.98 | 2.50 | 2.05 | 60 | 51 | 49 |
| 17 | CB | ND | 1500 | 3.40 | 3.08 | 2.75 | 70 | 63 | 61 |
| 18 | CB | ND | 3000 | 3.08 | 2.58 | 2.13 | 63 | 56 | 55 |
| 19 | CB | PW | 1500 | 3.31 | 2.98 | 2.63 | 80 | 76 | 76 |
| 20 | CB | PW | 3000 | 3.08 | 2.53 | 2.09 | 65 | 59 | 57 |
| 21 | CB | EB | 1500 | 3.47 | 3.28 | 3.04 | 82 | 79 | 79 |
| 22 | CB | EB | 3000 | 3.12 | 2.67 | 2.27 | 67 | 61 | 60 |
| 23 | CB | FII | 1500 | 3.45 | 3.23 | 2.97 | 77 | 73 | 72 |
| 24 | CB | FII | 3000 | 3.08 | 2.53 | 2.09 | 66 | 60 | 58 |